\documentclass[
    ,final            % use final for the camera ready runs
    ,sort&compress    % add further options here if necessary
  ]
  {aipproc}

\layoutstyle{6x9}
   \usepackage{graphicx}
  \usepackage{amssymb}

\def\teq#1{$\, #1\,$}                         % text equation
\def\apj{ApJ}

\def\asr{Adv. Space Res.}                       % DO NOT DELETE
\def\jgr{J. Geophys. Res.}

\def\ssr{Space Sci. Rev.}                       % DO NOT DELETE
\newcommand{\vol}[2]{$\,$\rm #1\rm , #2.}           
\begin{document}

\begin{flushright}
\phantom{p}
\vspace{-40pt}
	To appear in Proc. 4th IGPP International Astrophysics Conference,\\
	eds. G. Li, G. Zank and C. Russell (AIP Conf. Proc., New York)
\end{flushright} 

\title{Diffusive Acceleration of Ions\\ at Interplanetary Shocks}

\author{Matthew G. Baring \& Errol J. Summerlin}{
    address={Department of Physics and Astronomy, MS-108,
                      Rice University, P. O. Box 1892, \\
                      Houston, TX 77251-1892, USA\\
                      {\rm Email: baring@rice.edu, xerex@rice.edu}}
}

\begin{abstract}
Heliospheric shocks are excellent systems for testing theories of
particle acceleration in their environs.  These generally fall into two
classes: (1) interplanetary shocks that are linear in their ion
acceleration characteristics, with the non-thermal ions serving as test
particles, and (2) non-linear systems such as the Earth's bow shock and
the solar wind termination shock, where the accelerated ions strongly
influence the magnetohydrodynamic structure of the shock.  This paper
explores the modelling of diffusive acceleration at a particular
interplanetary shock, with an emphasis on explaining in situ
measurements of ion distribution functions.  The observational data for
this event was acquired on day 292 of 1991 by the Ulysses mission.  The
modeling is performed using a well-known kinetic Monte Carlo simulation,
which has yielded good agreement with observations at several
heliospheric shocks, as have other theoretical techniques, namely hybrid
plasma simulations, and numerical solution of the diffusion-convection
equation.  In this theory/data comparison, it is demonstrated that
diffusive acceleration theory can, to first order, successfully account
for both the proton distribution data near the shock, and the observation
of energetic protons farther upstream of this interplanetary shock than
lower energy pick-up protons, using a single turbulence parameter.   The
principal conclusion is that diffusive acceleration of inflowing
upstream ions can model this pick-up ion-rich event without the invoking
any seed pre-acceleration mechanism, though this investigation does not
rule out the action of such pre-acceleration.  
\end{abstract}

\maketitle

\section{Introduction}
\label{sec:Introduction}

Evidence for efficient particle acceleration at collisionless shocks in
the heliosphere abounds, including direct measurements of accelerated
populations in various energy ranges at the Earth's bow shock (e.g.
\cite{Schol80,Gosl89}) and interplanetary shocks (for the pre-Ulysses
era see, for example, \cite{SvA74,DPK81,Tan88}).  The development of
theories of shock acceleration is therefore strongly motivated, and a
variety of approaches have emerged.  One possible means for the
generation of non-thermal particles is the Fermi mechanism, often called
diffusive shock acceleration; this process forms the focus of this
paper.

There are various approaches to modelling diffusive shock acceleration.
Among these are hybrid and full plasma codes (e.g.
\cite{Quest88,Winske90,GBS92,KS95}), which place an emphasis primarily
on plasma structure and wave properties in the environs of shocks, and
the convection-diffusion differential equation approach \cite{KangJ95}.
In addition, the kinematic Monte Carlo technique of Ellison and Jones
(e.g., \cite{EJE81,JE91,EBJ95}) also focuses on diffusion and
convection, and describes the injection and acceleration of particles
from thermal energies to the highest relevant energies, addressing both
spectral and hydrodynamic properties.  The simulation technique makes no
distinction between accelerated particles and thermal ones, using an
identical phenomenological description of diffusion for both.  In this
work, following previous invocations, it is assumed that a particle's
mean free path \teq{\lambda} is proportional to its gyroradius
\teq{r_g}, i.e. \teq{\lambda =\eta r_g}, with \teq{\eta =}const. for all
particle momenta.  Upstream plasma quantities are input from
observational data, and downstream quantities are determined using the
full MHD Rankine-Hugoniot relations.

The Monte Carlo technique was used by Ellison et al. \cite{EMP90} to
perform the first detailed theory/data comparison for the quasi-parallel
portion of the Earth's bow shock.  They compared predictions of the
Monte Carlo method with particle distributions of protons, \teq{He^{++}} and
a $C$, $N$ and $O$ ion mix obtained by the AMPTE experiment.  The agreement
between model predictions and data was impressive, but required modeling
in the non-linear acceleration regime, when the dynamic effects of the
accelerated particles control the shock structure.  A similar
theory/data comparison was explored for interplanetary (IP) shocks in
the work of Baring et al. \cite{BOEF97}, where impressive agreement was
found between the Monte Carlo predictions and spectral data obtained by
the Solar Wind Ion Composition Spectrometer (SWICS) aboard Ulysses, in
the case of two shocks observed early in the Ulysses mission.  Such
agreement was possible only with the assumption of strong particle
scattering (i.e. near the Bohm diffusion limit) in the highly oblique
candidate shocks.  For a third shock, detected a month later, the
comparison failed with significant differences arising in the 500-800
km/sec range of the phase space distribution. Baring et al.
\cite{BOEF97} attributed this discrepancy to the omission of pick-up
ions from the model: such an extra component would be expected to
provide a substantial contribution to the accelerated population in this
particular event.

This paper explores the role of pick-up ions in such shocks via modeling
the accelerated population for the specific IP shock detected by the
SWICS and HI-SCALE instruments aboard the Ulysses spacecraft at around
4.5 AU, as reported in \cite{Gloeck94}.  Phase space distributions from
the simulations are compared with SWICS and HI-SCALE data, yielding
acceptable fits for the proton populations using standard prescriptions
for the injected pick-up ion distribution.  The simulation results
successfully account for the observation of energetic protons farther
upstream of the forward shock than lower energy pick-up protons, since a
rigidity-dependent diffusion is used in the modeling.

\section{The Ulysses Event of Day 292, 1991}
 \label{sec:model}

The forward shock of a CIR encountered by Ulysses on Day 292 of 1991, is
appropriate for a case study, with downstream particle distributions
published in Gloeckler et al. \cite{Gloeck94}. Various plasma parameters
for this shock were input for the Monte Carlo simulation, and were
obtained from \cite{Gloeck94} and the data compilations of
\cite{Balogh95,Hoang95}.   The shock was quite oblique, with
\teq{\theta_{Bn1} = 50^{\circ}\pm 11^{\circ}} being the angle the
upstream magnetic field made with the shock normal. It was also quite
weak, with a sonic Mach number of  \teq{M_{\rm s}\sim 2.53}, and
\cite{Gloeck94} inferred a value of \teq{r=u_1/u_2=2.4 \pm 0.3} for the
velocity compression ratio. The normalization of solar wind proton
distributions was established using \teq{n_p=2.0}cm$^{-3}$ as the solar
wind proton density.  Other parameters, such as the fluid speeds and
upstream plasma temperatures, are detailed in Summerlin \& Baring
\cite{SB05}, yielding an upstream flow speed of \teq{u_{1} \approx 55}
km/s in the shock rest frame. The pick-up proton distribution input for
the Monte Carlo simulation was taken from \cite{EJB99}, a developed
expression that is modeled on the seminal work of \cite{VS76}, and is 
similar in conception to pick-up ion distributions used in \cite{leRoux96}. The
pick-up ion model provides both the detailed shape and normalization of
this superthermal distribution at 4.5AU; it incorporates the gravitational 
focusing of interstellar neutrals, the physics of their ionization as a function 
of distance from the sun, and adiabatic losses incurred during propagation 
away from the sun.

The Monte Carlo shock acceleration simulation is described in
\cite{JE91,EBJ95,BOEF97,SB05}.  Particles are injected upstream and
allowed to convect into the shock, meanwhile diffusing in space so as to
effect multiple shock crossings, and thereby gain energy through the shock
drift and Fermi processes.  The particles gyrate in laminar
electromagnetic fields, with their trajectories being obtained by solving  
the Lorentz force equation in the shock rest frame, in which there is, in 
general, a {\bf u $\times$ B} electric field in addition to the magnetic
field. The effects of magnetic turbulence are modeled by scattering these
ions in the rest frame of the local fluid flow. While the simulation can
routinely model either large-angle or small-angle scattering, in this   
paper, large-angle scattering is employed, appropriate for the  
turbulent fields in IP shocks.  For this phenomenological scattering, it  
is assumed that a particle's mean free path \teq{\lambda} is proportional
to its gyroradius \teq{r_g}, i.e. \teq{\lambda =\eta r_g}, with \teq{\eta
=}const. for all particle momenta.  Other dependences on particle rigidity
can be employed, however the results are not extremely sensitive to such
choices.

At every scattering, the direction of the particle's momentum vector is
randomized in the local fluid frame, with the resulting effect that the
gyrocenter of a particle is shifted randomly by a distance of the order of
one gyroradius in the plane orthogonal to the local field.  Accordingly,
cross-field diffusion emerges naturally from the simulation, and is
governed by a kinetic theory description \cite{FJO74,EBJ95}, where the
ratio of the spatial diffusion coefficients parallel
(\teq{\kappa_\parallel =\lambda v/3}) and perpendicular
(\teq{\kappa_\perp}) to the mean magnetic field is given by
\teq{\kappa_\perp /\kappa_\parallel = 1/(1+\eta^2)}.  Clearly then,
\teq{\eta} couples directly to the amount of cross-field diffusion, and is
a measure not only of the frequency of collisions between particles and
waves, but also of the level of turbulence present in the system, i.e. is
an indicator of \teq{\langle \delta B/B\rangle}.  The Bohm diffusion limit
of quasi-isotropic diffusion, presumably corresponding to \teq{\langle
\delta B/B\rangle\sim 1}, is realized when \teq{\eta\sim 1}. As will
become apparent, \teq{\eta} is a parameter that critically controls the
injection efficiency of low energy particles, and the upstream diffusion
scale of accelerated ions.  The simulation outputs particle fluxes and
phase space distributions at any location upstream or downstream of the
shock, and in any reference frame including that of the Ulysses
spacecraft.  This capability renders it ideal for comparison with
observational data.

% FIGURE 1 GOES HERE

\begin{figure}
 \centerline{
  \includegraphics[width=.51\textwidth]{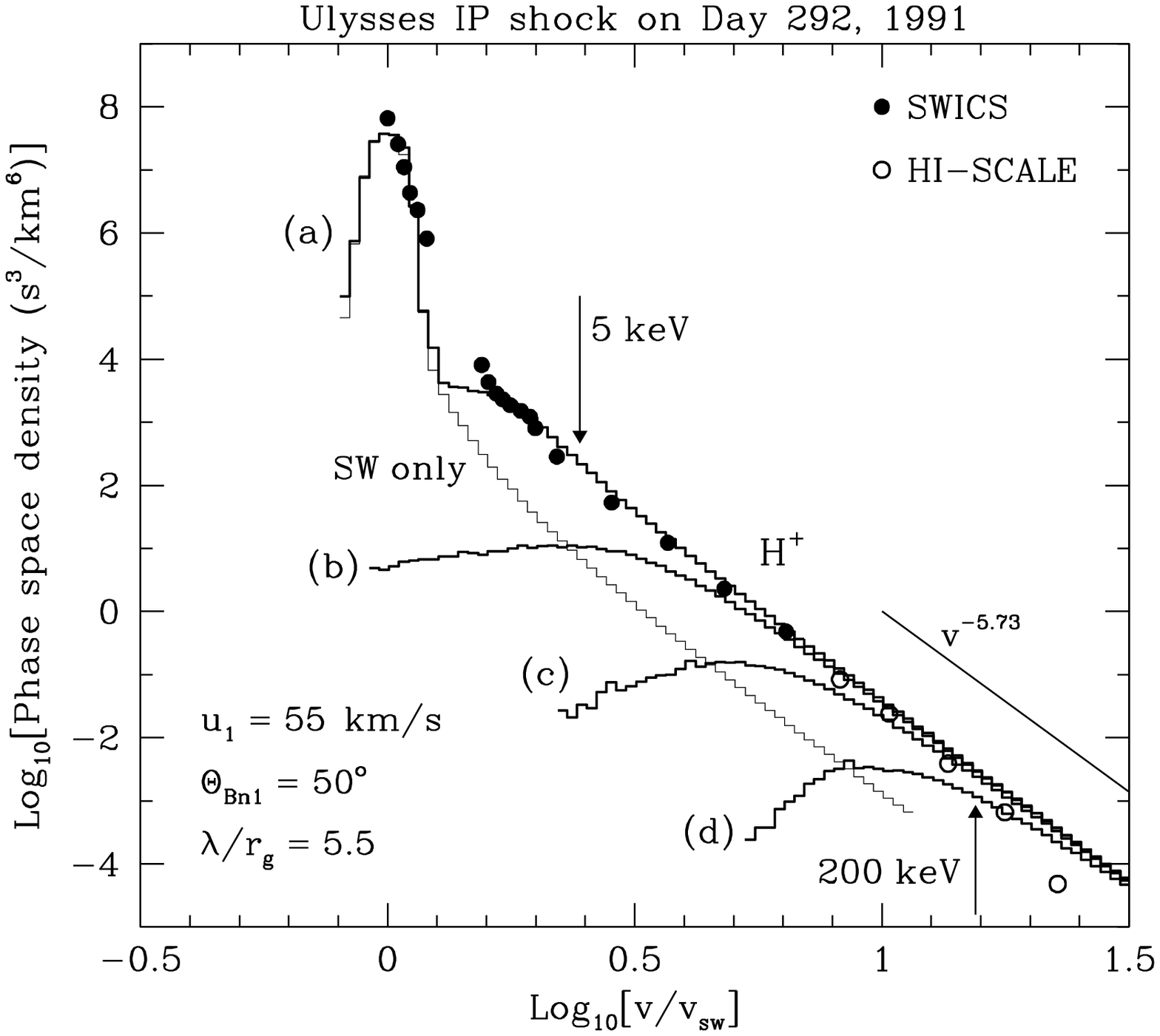}
   \hskip -0.1truecm
    \includegraphics[width=.51\textwidth]{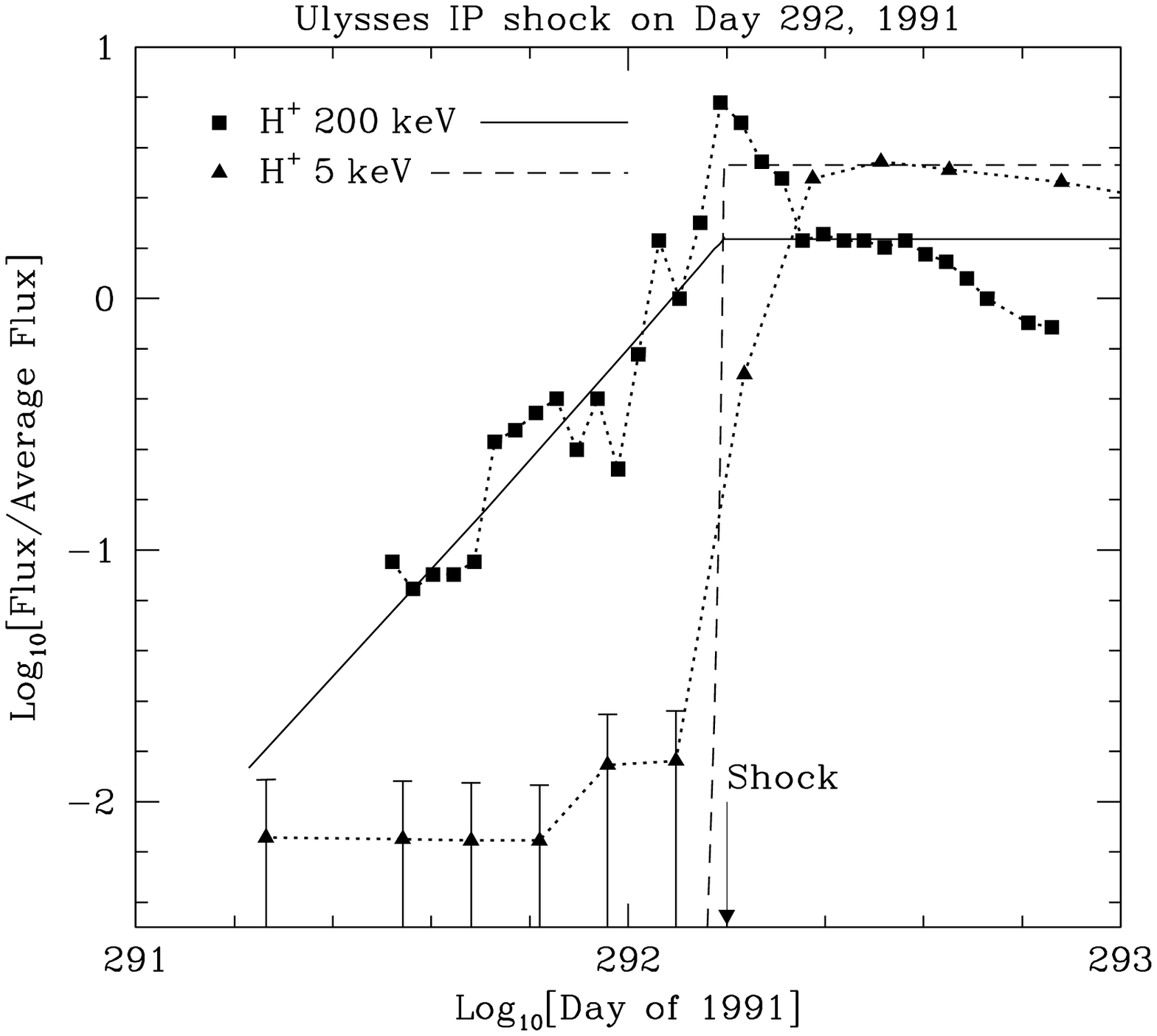}
  }
  \caption{{\it Left panel:} Comparison between phase space velocity 
distribution functions for data collected by the Ulysses mission for the
shock on day 292 of 1991, and Monte Carlo model results.  The data are
for \teq{H^+} (filled circles for SWICS data; open circles for HI-SCALE
points) solar wind and pickup ions, and are taken from Gloeckler et al.
\cite{Gloeck94}.  The heavyweight histograms are the corresponding Monte
Carlo models of acceleration of protons for \teq{u_1=55}km/sec, using
the optimal choice of plasma shock parameters from \cite{Gloeck94} and
sources indicated in the text (see also \cite{SB05}).  These four
spectra correspond to (a) downstream, and successively increasing times
upstream of the shock encounter, i.e. (b) 14 minutes, (c) 69 minutes and
(d) 278 minutes upstream.  The lighter weight histogram marked ``SW
only'' was for a run where pick-up ions were omitted.  The velocity axis
is the ratio of the ion speed \teq{v}, as measured in the spacecraft
frame, to the solar wind speed.  The model assumed \teq{\eta =\lambda
/r_g=5.5} and a shock of compression ratio \teq{r=2.1}, corresponding to
diffusive acceleration power-laws of index \teq{-5.73}, is indicated.
\newline
{\it Right panel:} The flux variations of accelerated pick-up ion
populations as a function of time near the shock.  The data for 5 keV
and 200 keV pick-up \teq{H^+} are depicted by filled triangles and
squares, respectively, and are taken from \cite{Gloeck94}. The Monte
Carlo model generated fluxes at different distances normal to the shock,
and were converted to spacecraft times by incorporating solar wind
convection.  The 5 keV and 200 keV pick-up \teq{H^+} traces are
displayed as dashed and solid curves, respectively, and exhibit an
exponential decline upstream of the shock that is characteristic of
diffusive shock acceleration.
}
\end{figure}

Figure~1 displays downstream distributions for thermal, pick-up and
accelerated protons from the Monte Carlo simulation and the SWICS and
HI-SCALE measurements (see Fig.~1 of \cite{Gloeck94}) taken in the frame
of the spacecraft on the downstream side of the Day 292, 1991 shock. 
The solar wind and pick-up proton parameters are fairly tightly
specified, so that the model has one largely free parameter, the ratio
of the particle mean free path to its gyroradius, \teq{\eta =\lambda
/r_g}. The efficiency of acceleration of thermal ions in oblique shocks,
i.e. the normalization of the non-thermal power-law, is sensitive
\cite{EBJ95,BOEF97} to the choice of \teq{\eta}, so this parameter was
adjusted to obtain a reasonable ``fit'' to the data. Here, the
accelerated pick-up ion phase space density is about a factor of 30
greater than that of the solar wind ions, denoted in the figure by the
``SW only'' histogram.

The downstream fit in the left hand panel of Fig.~1 models the accelerated
protons well, for \teq{\eta = 5.5\pm 1.5}, a value that is slightly higher than
those inferred in the fits of \cite{BOEF97} for shocks at around 2--3 AU,
yet consistent with a moderate level of field turbulence.  The uncertainty
in the inferred value of \teq{\eta} is due mostly to the observational 
uncertainty in the shock obliquity
\teq{\theta_{Bn1}}. The non-thermal proton distribution is composed
virtually entirely of accelerated pick-up ions: the accelerated thermal
\teq{H^{+}} ions are injected much less efficiently in the simulation than
in the observations. The efficiency of acceleration of thermal ions could
be increased via several means: (i) by lowering the shock obliquity angle
\teq{\theta_{Bn1}}, for which there is a large observational uncertainty;
(ii) by decreasing \teq{\eta}, corresponding to increased turbulence,   
without altering the pick-up ion acceleration efficiencies substantially,
and (iii)  increasing the temperature of the thermal ions somewhat, though
this would reduce the compression ratio and accordingly steepen the
non-thermal continuum.  Note that the distribution of accelerated
\teq{He^+} pick-up ions reported by \cite{Gloeck94} for this shock can be
modeled by the {\it same} scattering parameter \teq{\eta =5.5}.  This   
enticing property is addressed by Summerlin \& Baring \cite{SB05}, where
it is observed that the \teq{He^{++}} distribution requires lower
\teq{\eta} for a fit of comparable quality.

An instructive diagnostic on the acceleration model is to probe the
spatial scale of diffusion.  This is routinely performed with the Monte
Carlo simulation by placing flux measurement planes upstream of the shock
at different distances, as well as downstream. Results are illustrated in
the left hand panel of Fig.~1 via the display of upstream distributions of
high energy particles at different times, i.e. distances from the shock.
The Figure exhibits the characteristic ``peel-off'' effect where 
superthermal ions become depleted at successively high energies the
further the detection plane is upstream of the shock; this signature was
first identified by Lee \cite{Lee82}.  Gloeckler et al. \cite{Gloeck94}
discussed an energy-dependent rise in fluxes of non-thermal particles {\it
prior} to the shock crossing. This was cited as indicating the existence
of a pre-acceleration mechanism. Fluxes for two different \teq{H^+} ion
energies, 5 keV and 200 keV, were obtained from spectra like those in the
left hand panel in the Fig.~1, and are displayed in the right hand panel
of the Figure, together with corresponding data from Fig.~3 of
\cite{Gloeck94} for identical energy windows.  Note that the Ulysses data
normalization was established by averaging over 3 days of data, whereas
the model normalization was adjusted to match observed fluxes around 1/2
day downstream of the shock.

It is clear that the spatial scale of the exponential decline of ions
upstream of the shock is more or less identical to that of the model,
for our choice of \teq{\eta=5.5}. High energy particles with a mean free
path \teq{\lambda\propto r_g} establish an exponential dilution in
space/time due to random scattering of the particles as they leak
upstream against a convective flow. For the 200 keV ions with their
relatively long mean free paths, the simulation results are clearly well
correlated with the data prior to the shock, modulo plasma fluctuations,
and in particular the overshoot just downstream. On the other hand, for
the lower energy 5keV ions, the exponential decay has a very short time
scale, around a factor of 40 smaller than for the 200 keV ions,
realizing background levels upstream until very close to the shock.  So,
although the simulation results are consistent with the observed
results, it is impossible to draw more definitive conclusions without an
improvement in data time resolution, or a focus on ions of intermediate
energy, say around 50 keV. Note that while this comparison is
suggestive, it does not conclusively prove that diffusion is the
dominant operating mechanism in this system. Yet alternative
explanations must generate exponential declines that are consistent with
convective loss scales of the order of a few gyroradii, with the
physical mechanism responsible for transport upstream being also a
direct cause of injection into the acceleration process.

\section{Conclusions}
\label{sec:conclusion}

This paper has compared the phase space distributions for protons from
the Monte Carlo simulation of diffusive shock acceleration with those
observed by the Ulysses instruments SWICS and HI-SCALE in the Day 292,
1991 shock.  There is a good deal of consistency between theory and
experiment for the energetic protons above speeds around 600 km/sec, an
agreement that is extended to include \teq{He^+} pick-up ion spectra in
\cite{SB05}.  At these speeds, the injection of pick-up protons
dominates that of solar wind protons. The normalization of the energetic
proton power-law establishes \teq{\eta = 5.5}, where \teq{\lambda =\eta
r_g} is the diffusive mean free path.  This provides substantial
cross-field diffusion (\teq{\kappa_\perp /\kappa_\parallel \approx 0.03}), 
the prerequisite for efficient injection and acceleration in this diffusive 
model, when highly oblique shocks are being simulated.

The upstream spatial scales of the acceleration were also probed, with
the flux increases of energetic protons seen upstream of the shock being
well-modeled by the expected upstream ``leakage'' associated with
diffusive acceleration.  The value of \teq{\eta =5.5} inferred from the
spectral fit scales the upstream diffusive lengthscale, and the
accompanying exponential decline in predicted flux is commensurate with
the Ulysses data presented in Gloeckler et al. \cite{Gloeck94}.  Hence,
the observed upstream flux precursor is not clear evidence of a
pre-acceleration mechanism, as claimed by \cite{Gloeck94}, though it is
quite possible that some pre-acceleration mechanism may be acting. The
flux fluctuations in time clearly indicate the contribution of a
non-diffusive process in the plasma shock, effects that are not
incorporated in the simulation. Yet, the fact that the diffusive model
works so well in coupling the spectral and spatial properties suggests
that diffusion is an integral part of the acceleration process at this
shock.

\bibliographystyle{aipproc}

\end{document}